\documentclass[12pt,a4paper]{article}

\usepackage[utf8]{inputenc}
\usepackage[T1]{fontenc}
\usepackage{lmodern}            
\usepackage{microtype}          
\usepackage{geometry}           
\usepackage{parskip}            
\usepackage{setspace}           
\usepackage{xcolor}             
\usepackage{hyperref}           
\usepackage{fancyhdr}           
\usepackage{titlesec}           
\usepackage{booktabs}           
\usepackage{tabularx}           
\usepackage{array}              
\usepackage{multirow}           
\usepackage{amsmath}            
\usepackage{amssymb}            
\usepackage{amsthm}             
\usepackage{enumitem}           
\usepackage{graphicx}           
\usepackage{float}              
\usepackage[round]{natbib}      

\definecolor{linkblue}{RGB}{0,102,204}
\definecolor{citegreen}{RGB}{0,153,0}
\definecolor{urlred}{RGB}{204,0,0}
\definecolor{sectioncolor}{RGB}{0,51,102}

\hypersetup{
    colorlinks=true,
    linkcolor=linkblue,
    urlcolor=urlred,
    citecolor=citegreen,
    pdftitle={Necessary-But-Not-Sufficient Benchmarks for AI Reasoning: Bhatt Conjectures},
    pdfauthor={Manish Bhatt},
    pdfkeywords={AI reasoning, understanding, benchmarks, tautologies, deep learning},
    pdfsubject={AI capability benchmarks}
}

\titleformat{\section}
    {\Large\bfseries\sffamily\color{sectioncolor}}
    {\thesection}
    {1em}
    {}
    [\vspace{-0.5em}]
    
\titleformat{\subsection}
    {\large\bfseries\sffamily}
    {\thesubsection}
    {1em}
    {}
    [\vspace{-0.3em}]

\newcommand{\TU}{\ensuremath{\mathbf{T_U}}}
\newcommand{\TUstar}{\ensuremath{\mathbf{T_U^{*}}}}
\newcommand{\Tone}{\ensuremath{\mathbf{T_1}}}

\theoremstyle{definition}
\newtheorem{definition}{Definition}[section]

\title{%
    \vspace{-2em}
    \Large\textsc{Bhatt Conjectures: On Necessary-But-Not-Sufficient Benchmark Tautology for Human Like Reasoning}
}
\author{%
    \Large Manish Bhatt\footnote{This work is not related to author's position at Amazon.}\\
    \normalsize\textit{AI Security Researcher}\\
    \normalsize\href{mailto:manish.bhatt13212@gmail.com}{manish.bhatt13212@gmail.com}
}
\date{\large June 2025}

\begin{document}
\maketitle
\thispagestyle{fancy}

\section{Introduction}

Debates about whether Large Language/Reasoning Models (LLMs/LRMs) truly \emph{reason} or merely \emph{pattern-match} \citep{marcus2023reasoning, chollet2019measure} suffer from shifting goal-posts. In my opinion, two analytic—hence ``tautological''—benchmarks cut through that fog in my mental model:

\begin{itemize}[label=\textbullet, itemsep=0.5em]
    \item \textbf{\Tone}: \emph{Reasoning-Capability Tautology}
    \item \textbf{\TU}: \emph{Understanding-Capability Tautology}
\end{itemize}

Because each definition states \emph{necessary and sufficient} conditions \textbf{by construction}, the challenge is \textbf{not} to ``prove the tautologies are true''—they are definitions—but to determine whether a \emph{concrete system} meets them. 

This paper first restates \Tone{} and \TU{} with surgical precision, then extends \TU{} to a richer \TUstar{} that incorporates causal modelling \citep{pearl2009causality}, metacognition \citep{flavell1979metacognition}, the fast/slow thinking dichotomy \citep{kahneman2011thinking}, and the (currently untestable) question of phenomenal awareness \citep{baars1988cognitive}.

\section{\texorpdfstring{\Tone}{T1} — Reasoning-Capability Tautology}

\begin{definition}[\Tone]
A system \textbf{reasons} with respect to a problem class $C$ \textbf{if and only if}:
\begin{enumerate}[label=\textbf{(R\arabic*)}, itemsep=0.5em]
    \item it produces a correct solution from \emph{any} logically equivalent representation of the formally specified premises, and
    \item its success probability remains high when the surface form of the problem is \emph{outside} its training distribution (no isomorphic instance or step-by-step demonstration seen in training or finetuning data).
\end{enumerate}
\end{definition}

\subsection{Canonical Corollaries}

\begin{table}[H]
\centering
\renewcommand{\arraystretch}{1.4}
\begin{tabularx}{\textwidth}{@{}l>{\raggedright\arraybackslash}X>{\raggedright\arraybackslash}X@{}}
\toprule
\textbf{ID} & \textbf{Corollary} & \textbf{Example Diagnostic Test} \\
\midrule
\textbf{C1} & \textbf{Representation Invariance}\\
& Restate the task in a different natural language, formal notation, diagram, or noisy encoding; accuracy must persist. 
& Paraphrase a Sudoku constraint set into first-order logic; solver must still succeed. \\
\addlinespace[0.5em]
\textbf{C2} & \textbf{Complexity Scaling}\\
& Hold logical structure fixed, increase instance size (e.g., more discs in Tower of Hanoi). Accuracy should not collapse. 
& Apple LRMs drop to $\approx 0\%$ beyond 5 discs despite flawless small-$n$ performance \citep{shojaee2025apple}. \\
\addlinespace[0.5em]
\textbf{C3} & \textbf{Zero-Shot Robustness}\\
& Inject \emph{novel} but logically equivalent surface patterns absent from training. 
& Swap ``$A\to B$'' syntax for $\lambda$-calculus encoding; solution quality must persist. \\
\bottomrule
\end{tabularx}
\caption{Corollaries and diagnostic probes for \Tone.}
\label{tab:t1-corollaries}
\end{table}

\section{\texorpdfstring{\TU}{T\_U} — Understanding-Capability Tautology}

\begin{definition}[\TU]
A system \textbf{understands} a knowledge domain $D$ \textbf{if and only if} for every proposition $\varphi$ about $D$:
\begin{enumerate}[label=\textbf{(U\arabic*)}, itemsep=0.5em]
    \item it maps \emph{any} truth-preserving representation $\rho(\varphi)$ to an internal state $I(\varphi)$ that recovers the truth value of $\varphi$, and
    \item conditions U1 hold when $\rho(\varphi)$ stems from a distribution statistically independent of the training data, and when queried with \emph{unseen} operations derivable solely from $I(\varphi)$ (e.g., counterfactuals, entailments).
\end{enumerate}
\end{definition}

\subsection{Canonical Corollaries}

\begin{table}[H]
\centering
\renewcommand{\arraystretch}{1.4}
\begin{tabularx}{\textwidth}{@{}l>{\raggedright\arraybackslash}X>{\raggedright\arraybackslash}X@{}}
\toprule
\textbf{ID} & \textbf{Corollary} & \textbf{Example Diagnostic Test} \\
\midrule
\textbf{C4} & \textbf{Modal Invariance}\\
& Understanding survives cross-modal transfer (text $\rightarrow$ image schema, Braille, speech). 
& Encode ``water freezes at 0°C'' as a line in a phase diagram; model must still answer ``True/False?'' \\
\addlinespace[0.5em]
\textbf{C5} & \textbf{Counterfactual Competence}\\
& Given $I(\varphi)$, the system answers \emph{derived} queries (inference, contradiction, analogy). 
& From ``All swans are birds,'' infer ``If $X$ is a swan, $X$ can fly \emph{unless} flightless-bird exceptions apply.'' \\
\addlinespace[0.5em]
\textbf{C6} & \textbf{Distribution Shift}\\
& Replace everyday examples with rare or synthetic ones; truth evaluations remain stable. 
& Test chemical facts using IUPAC strings never seen during training. \\
\bottomrule
\end{tabularx}
\caption{Corollaries and diagnostic probes for \TU.}
\label{tab:tu-corollaries}
\end{table}

\subsection{Relation to \texorpdfstring{\Tone}{T1}}

\TU{} \emph{strictly contains} \Tone{} on factual domains: correct reasoning over $\varphi$ demands first mapping its meaning robustly; hence $\TU \Rightarrow \Tone$ (but not conversely).

\section{\texorpdfstring{\TUstar}{T\_U*} — Extended Understanding}

\TU{} is intentionally behaviourist. To bridge explanatory gaps without leaping to metaphysics, we propose an \emph{Extended Understanding-Capability Tautology}, \TUstar:

\begin{definition}[\TUstar]
A system \textbf{deeply understands} a domain $D$ \textbf{if and only if} it satisfies \TU{} \emph{and} the following premises:
\begin{enumerate}[label=\textbf{(E\arabic*)}, itemsep=0.5em]
    \item \textbf{Causal Structural Fidelity} — Its internal representation mirrors the \emph{causal} graph of $D$, enabling do-calculus interventions \citep{pearl2009causality}.
    \item \textbf{Metacognitive Self-Awareness} — It evaluates and calibrates the reliability of its own inferences, signalling uncertainty or ignorance \citep{shum2023metacognitive}.
    \item \textbf{Phenomenal Awareness} — Inference is accompanied by subjective experience (qualia). Untestable, this marks the boundary between functional mastery and conscious sapience \citep{chalmers1995facing}.
\end{enumerate}
\end{definition}

\section{Fast \& Slow Thinking: A Path to Emergent Cognitive Abilities}

\paragraph{Open Challenge.} 
Integrating both modes is non-trivial. Early attempts (e.g., hypothetical models like \emph{Claude 3.7}, \emph{Qwen 3.2}) interleave chain-of-thought prompting \citep{wei2022chain} with high-temperature sampling yet still default to shallow heuristics. Robust solutions may require:

\begin{itemize}[itemsep=0.5em]
    \item \textbf{Hybrid architectures} that switch between fast pattern recall and slow causal reasoning based on task uncertainty \citep{suzgun2023challenging}.
    \item \textbf{On-device ``fast cores''} (e.g., future \emph{Llama} models) tightly coupled with stronger cloud ``slow cores'' (e.g., a future \emph{Claude} model), delivering low-latency intuition plus high-fidelity deliberation.
    \item \textbf{Curriculum training} rewarding extended, coherent reasoning chains while penalising hallucination \citep{lightman2023lets}.
\end{itemize}

\subsection{Layered Evidence Framework}

\begin{table}[H]
\centering
\renewcommand{\arraystretch}{1.4}
\begin{tabularx}{\textwidth}{@{}>{\raggedright\arraybackslash}p{3cm}>{\raggedright\arraybackslash}X>{\raggedright\arraybackslash}X@{}}
\toprule
\textbf{Layer} & \textbf{Target Property} & \textbf{Test Technology} \\
\midrule
\textbf{Behavioural} (\TU) 
& Format- \& distribution-robust truth preservation 
& Stress tests; adversarial modalities \citep{hendrycks2021measuring} \\
\addlinespace[0.5em]
\textbf{Causal} (E1) 
& Fidelity to real-world causal graph 
& Interventional datasets; SCM extraction; do-calculus probes \citep{xia2021causal} \\
\addlinespace[0.5em]
\textbf{Metacognitive} (E2) 
& Calibrated self-uncertainty 
& Selective prediction; verifier LLMs; abstention metrics \citep{kadavath2022language} \\
\addlinespace[0.5em]
\textbf{Fast/Slow Arbitration} 
& Dynamic mode switching 
& Latency studies; scratchpad audits; reasoning-depth benchmarks \\
\addlinespace[0.5em]
\textbf{Phenomenal} (E3) 
& Qualia / ``something-it-is-like'' 
& \emph{Open research problem}—no agreed operationalisation \citep{nagel1974what} \\
\bottomrule
\end{tabularx}
\caption{Evidence stack incorporating fast/slow thinking.}
\label{tab:evidence-stack}
\end{table}

\section{Necessity \& Sufficiency Summary}

\begin{table}[H]
\centering
\renewcommand{\arraystretch}{1.3}
\resizebox{\textwidth}{!}{%
\begin{tabular}{@{}l@{\hspace{1em}}c@{\hspace{1em}}c@{\hspace{1em}}c@{\hspace{1em}}c@{\hspace{1.5em}}p{4.8cm}@{}}
\toprule
\textbf{Property} & \textbf{\Tone} & \textbf{\TU} & \textbf{\TUstar} & \textbf{Fast/Slow} & \textbf{Still Missing} \\
\midrule
Correct logical inference       & \checkmark & \checkmark & \checkmark & \checkmark & — \\
Truth-robust representation     & —          & \checkmark & \checkmark & \checkmark & — \\
Causal structural fidelity      & —          & —          & \checkmark & —          & Interventional benchmarks \\
Metacognitive awareness         & —          & —          & \checkmark & \checkmark & Self-calibration metrics \\
Phenomenal experience          & —          & —          & - & ?          & ? \\
\bottomrule
\end{tabular}%
}
\caption{Properties satisfied by each benchmark level.}
\label{tab:properties-summary}
\end{table}

\section{Limitations and Scope}

This paper's benchmarks are definitional, not exhaustive. We explicitly acknowledge several critical dimensions that fall outside the scope of this framework:

\begin{itemize}[label=\textbullet, leftmargin=*, itemsep=0.5em]
    \item \textbf{Learning and Development.} This framework defines the \emph{what} (the criteria for understanding) but does not prescribe the \emph{how} (the mechanisms for achieving it). It omits a discussion of developmental learning theories \citep{gopnik2009philosophical} or the specific architectural and curriculum requirements needed to progress from \Tone{} to \TUstar.

    \item \textbf{Social and Multi-Agent Understanding.} The benchmarks evaluate an agent in cognitive isolation. They do not test for social reasoning, such as Theory of Mind \citep{premack1978does} (modeling the beliefs and intentions of others), collaborative problem-solving, or the intersubjective validation of knowledge.

    \item \textbf{Embodiment and Grounded Experience.} The framework is abstract and does not engage with the symbol grounding problem \citep{harnad1990symbol}. Whether the deepest levels of causal and conceptual understanding (E1) are possible without embodied interaction in a physical environment remains an open question not addressed here.

    \item \textbf{Methodological Feasibility.} While diagnostic tests are proposed, the paper does not fully resolve the practical hurdles of their implementation. Testing across all logically equivalent representations or reliably interpreting the internal causal state of a large neural network are major, unsolved research challenges in their own right \citep{lipton2018mythos}.
\end{itemize}

\section{Definitions and Terminology Clarifications}
\label{sec:definitions}

The discourse surrounding AI capabilities, particularly regarding "reasoning" and "understanding," is often complicated by a lack of precise definitions for these deeply loaded terms. To ensure clarity and avoid semantic ambiguities, this section explicitly delineates the specific interpretations of key terminology within the context of the Bhatt Conjectures. These definitions are operational, serving as the basis for the benchmarks proposed herein, and should not be conflated with broader philosophical or colloquial meanings unless explicitly stated.

\subsection{On the Nature of "Tautology"}

\begin{definition}[Tautology in this Context]
Within this paper's framework, a "tautology" refers to a **definition constructed analytically to state necessary and sufficient conditions for a specific AI capability (Reasoning, Understanding, or Deep Understanding) *as defined by the framework itself*.** Its truth is established by its definitional construction within this formal system.
\end{definition}

\noindent \textbf{Clarification:} These benchmarks are termed "tautological" because their validity *within the confines of this paper's framework* is inherent in their definitional construction. They rigorously delineate what we mean by "reasoning" and "understanding" for the purpose of evaluating AI systems against these precise criteria. This approach circumvents debates about the empirical "truth" of these definitions by asserting them as axiomatic for the evaluation model presented, rather than as empirically derived observations about intelligence in general.

\subsection{Defining "Reasoning" (\Tone)}

The term "reasoning" is frequently used to imply complex cognitive processes, logical inference, and abstract problem-solving. Our \Tone{} definition narrows this to empirically testable behavioral patterns:

\begin{definition}[\Tone{} - Operational Reasoning]
A system \textbf{reasons} with respect to a problem class $C$ \textbf{if and only if}:
\begin{enumerate}[label=\textbf{(R\arabic*)}, itemsep=0.5em]
    \item it produces a correct solution from \emph{any} logically equivalent representation of the formally specified premises, and
    \item its success probability remains high when the surface form of the problem is \emph{outside} its training distribution (no isomorphic instance or step-by-step demonstration seen in training or finetuning data).
\end{enumerate}
\end{definition}

\noindent \textbf{Clarification:} For an AI system to satisfy \Tone{}, it must demonstrate both **representation invariance** and **zero-shot robustness** in problem-solving. This implies an abstraction of the underlying logical structure that enables consistent application regardless of problem presentation (e.g., natural language, formal logic, diagrams, noisy encoding) and successful generalization to entirely novel instances. It explicitly excludes cases where performance relies solely on memorization or shallow pattern completion over familiar training data.

\subsection{Defining "Understanding" (\TU)}

"Understanding" colloquially suggests comprehension and the grasping of meaning beyond surface-level information. Our \TU{} definition builds upon \Tone{} by adding a robust representational requirement:

\begin{definition}[\TU{} - Operational Understanding]
A system \textbf{understands} a knowledge domain $D$ \textbf{if and only if} for every proposition $\varphi$ about $D$:
\begin{enumerate}[label=\textbf{(U\arabic*)}, itemsep=0.5em]
    \item it maps \emph{any} truth-preserving representation $\rho(\varphi)$ to an internal state $I(\varphi)$ that recovers the truth value of $\varphi$, and
    \item conditions U1 hold when $\rho(\varphi)$ stems from a distribution statistically independent of the training data, and when queried with \emph{unseen} operations derivable solely from $I(\varphi)$ (e.g., counterfactuals, entailments).
\end{enumerate}
\end{definition}

\noindent \textbf{Clarification:} \TU{} extends beyond \Tone{} by requiring **truth-robust representation**. An AI's 'understanding' of a domain necessitates the creation of an internal representation that faithfully preserves truth values across varied input modalities (e.g., text, image schemas, speech, formal encodings) and distributions. Critically, this internal representation must be sufficiently rich to support **counterfactual competence** and **novel inference** (e.g., deriving previously unseen logical entailments or contradictions), even when the queries themselves are statistically independent of the training data. This criterion implies a robust, abstract knowledge representation, as opposed to mere recall of facts or pattern extrapolation.

\subsection{Defining "Deep Understanding" (\TUstar)}

"Deep understanding" often carries connotations of profound comprehension, intuition, and sometimes even consciousness. \TUstar{} addresses specific facets of these intuitions while acknowledging inherent limits.

\begin{definition}[\TUstar{} - Operational Deep Understanding]
A system \textbf{deeply understands} a domain $D$ \textbf{if and only if} it satisfies \TU{} \emph{and} the following premises:
\begin{enumerate}[label=\textbf{(E\arabic*)}, itemsep=0.5em]
    \item \textbf{Causal Structural Fidelity} — Its internal representation mirrors the \emph{causal} graph of $D$, enabling do-calculus interventions \citep{pearl2009causality}.
    \item \textbf{Metacognitive Self-Awareness} — It evaluates and calibrates the reliability of its own inferences, signalling uncertainty or ignorance \citep{shum2023metacognitive}.
    \item \textbf{Phenomenal Awareness} — Inference is accompanied by subjective experience (qualia). Untestable, this marks the boundary between functional mastery and conscious sapience \citep{chalmers1995facing}.
\end{enumerate}
\end{definition}

\noindent \textbf{Clarification:} \TUstar{} captures capabilities beyond purely behavioral understanding. It mandates **causal structural fidelity**, meaning the system's internal knowledge representation must accurately reflect the true causal relationships within a domain, enabling it to correctly predict the outcomes of interventions (e.g., "what if I *did* X?"). It further requires **metacognitive self-awareness**, where the system can assess and self-calibrate the reliability of its own inferences, expressing uncertainty or ignorance in a principled manner. The inclusion of **phenomenal awareness** (subjective, qualitative experience) is explicitly designated as an untestable, philosophical boundary condition within \TUstar{}. This highlights the ultimate conceptual limit of 'deep understanding' without asserting its current achievability or empirical verifiability in artificial systems.

\subsection{Clarifying "Internal State"}

The concept of an "internal state" in AI can be vague, referring to anything from neuron activations to symbolic representations.

\begin{definition}[Internal State $I(\varphi)$]
An **internal state $I(\varphi)$** refers to the **specific, actionable computational representation** (e.g., a learned embedding, a pattern of activations, a symbolic data structure) generated by the AI system upon processing a proposition $\varphi$.
\end{definition}

\noindent \textbf{Clarification:} The existence of an internal state $I(\varphi)$ is inferred by its functional properties: its capacity to consistently and robustly "recover the truth value" of $\varphi$ across varied representations and to support novel, derived operations (as per \TU). This definition emphasizes functional computational representation and does not inherently imply human-like subjective experience or an easily interpretable symbolic form, though such forms might be a consequence of achieving the specified capabilities.

\section{Conclusion}

In my mental model, \Tone{} and \TU{} provide \emph{necessary but not sufficient} criteria for reasoning and understanding, respectively. The extended \TUstar{} adds causal and metacognitive layers, approaching—but not crossing—the phenomenal boundary. These tautologies offer somewhat of a rigorous framework for moving beyond ``Does it reason?'' to ``How deeply does it understand?''—a question whose answer will shape the trajectory of artificial general intelligence.

\section{Acknowledgments}

The author thanks colleagues in the AI research community for stimulating discussions on the nature of machine reasoning and understanding. Special appreciation goes to those who provided feedback on earlier drafts of these conjectures, helping to refine the definitions and clarify the conceptual boundaries. The author is grateful for the intellectual environment that encourages rigorous examination of foundational questions in artificial intelligence.

Special thanks to Jim Schwoebel for his feedback. 

\bibliographystyle{plainnat}


\end{document}